\documentclass[authoryear,preprint,review,12pt]{elsarticle}

\usepackage{amsmath}
\usepackage{amssymb}
\usepackage{graphicx}
\usepackage{natbib}
\usepackage{color}

\newcommand{\erf}[1]{\ensuremath{\mathrm{erf}(#1)}}
\newcommand{\nui}{\ensuremath{\tilde{\nu}_i}}
\newcommand{\dnu}{\ensuremath{\Delta\tilde{\nu}}}
\newcommand{\nuj}{\ensuremath{\tilde{\nu}_{0;j}}}
\newcommand{\cm}{\ensuremath{\mathrm{cm^{-1}}}}
\newcommand{\htwoo}{\ensuremath{\mathrm{H_2}\mathrm{^{16}O}}}

\journal{Icarus}

\begin{document}

\begin{frontmatter}



\title{Temperature-dependent molecular absorption cross sections for exoplanets and other atmospheres}


\author{Christian Hill, Sergei N. Yurchenko  and Jonathan Tennyson}

\address{Department of Physics and Astronomy, University College London, Gower Street, WC1E 6BT London, UK}

\begin{abstract}
  Exoplanets, and in particular hot ones such as hot Jupiters, require
  a very significant quantities of molecular spectroscopic data to
  model radiative transport in their atmospheres or to interpret their
  spectra. This data is commonly provided in the form of very
  extensive transition line lists.  The size of these line lists is
  such that constructing a single model may require the
  consideration of several billion lines. We present a procedure to
  simplify this process based on the use of cross sections. Line lists
  for water, H$_3^+$, HCN /HNC and ammonia have been turned into cross
  sections on a fine enough grid to preserve their spectroscopic
  features. Cross sections are provided at a fixed range of temperatures and
  an interpolation procedure which can be used to generate cross
  sections at arbitrary temperatures is described. A web-based
  interface (www.exomol.com/xsec) has been developed to allow
  astronomers to download cross sections at specified temperatures and
  spectral resolution. Specific examples are presented for the key
  water molecule. 
  

\end{abstract}


\end{frontmatter}


\section{Introduction}
\label{intro}

With the growing realization that many, probably most, stars support
exoplanets, developing the means to systematically characterize the
atmospheres of these planets has become a major scientific priority
\citep{jt523}. Given the likely complex chemistry of these
atmospheres and the elevated temperature that is found in the most
observable planets, there is a significant demand for spectroscopic
data on the probable exoplanet atmospheric constituents.

Recently we have launched a new project, called ExoMol (see
www.exomol.com), with the aim of providing molecular transition data
appropriate for exoplanet models which is reliable over a wide range
of temperatures \citep{jt529}. The ExoMol project involves
constructing line lists of spectroscopic transitions for key molecules
which are valid over the entire temperature and wavelength domain that
is likely to be astrophysically important for these species. Especially
for polyatomic molecules, these line lists can become very large with
hundreds of millions \citep{jt374,jt378,jt469,11TaPe.CO2} or even
billions \citep{jt500} of individual transitions needing to
characterized and stored.  A complete linelist for methane, for which
so far only a preliminary version is available \citep{09WaScSh.CH4},
can be expected to be even larger. Indeed potential line lists for
larger species, such as higher hydrocarbons, for which spectroscopic
data is needed for exoplanetary research, are likely to be so large 
as to potentially make their use impractical. 

Molecular line lists are being actively used to model the spectra of
exoplanets (eg \citet{jt495}) and cool brown dwarfs with similar
temperatures (eg \citet{jt484,ckg11}).  However, sampling billions of
individual transitions to model relatively low resolution astronomical
spectra is probably not necessary in many cases. An alternative
approach is to represent the molecular absorptions as cross sections
generated at an appropriate resolution and temperature. The advantage
of this approach is that the data handling issues related to dealing
with large data sets largely disappear. The disadvantage is that cross
sections are inflexible - a particular cross section set is only valid
for a single state of temperature and pressure. Cross sections are
therefore often regarded a second choice compared to maintaining a
full line list \citep{jt453}.

In this paper we develop a strategy whereby cross sections are
provided for the user in a flexible fashion without loss of accuracy
or the specificity of using a complete line list. To this end we have
provided a web application which, starting from very high resolution
cross sections generated for each molecule at a range of temperatures,
can provide cross sections at a temperature and resolution specified
by the user. Of course this approach is based on the implicit
assumption of local thermodynamic equilibrium (LTE) and any non-LTE
treatment will continue to have to rely on the explicit use of
transition line lists. So far, these cross sections do not consider
collisional broadening effects and are therefore, at their highest
resolution, appropriate for the zero pressure limit only.

The line lists for water \citep{jt378,jt469}, H$_3^+$ \citep{jt181, jt478}, HCN /HNC \citep{jt298, jt374, jt447} and ammonia \citep{jt500} 
were used to generate cross sections for these species.
For concreteness, this work uses the main water isotopologue, \htwoo, as its working
example.  Water is known to be a key species in exoplanetary atmospheres and the
BT2 line list has been used in studies of exoplanets
\citep{jt400,Swain09,tinetti10a,baraffe10,tinetti10b,Shabram11,Barman11,jt516}
as well in a large variety of planetary \citep{jt431,jt463,jeremy09},
astrophysical
\citep{Warren07,jt330,jt349,Burgasser08,jt452,jt417,jt357} and, indeed,
engineering \citep{Kr07,Lindermeir20121575} studies which generally
focus on the radiative transport by hot water.  The BT2 line list was
used as part of the recently updated HITEMP database \citep{jt480}. In
that work the size of the line list was reduced using a technique
based upon importance sampling at a range of key temperatures. In
practice the number of water lines in HITEMP remains large, over 100
million.

The calculation of opacities and other spectral properties
due to water vapour at these elevated temperatures can therefore
become onerous, and so we present here pre-calculated absorption cross
sections for a range of temperatures between 296~K and 3000~K, binned
to different resolutions. The highest resolution cross sections are
suitable for modelling low-density environments where only Doppler
broadening contributes to the line width whereas by binning to a
wavenumber grid spacing significantly larger than the
pressure-broadened half-width, higher-density environments are
described well by the calculated cross sections. However, no attempt
is made to include contributions to the opacity from the water vapour
continuum or water dimer absorption.

\section{Method}
The high-resolution cross section is calculated on an evenly-spaced wavenumber grid, $\nui$, defining bins of width $\dnu$. Only Doppler broadening is considered so each absorption line has a Gaussian shape:
\begin{align}
f_\mathrm{G}(\tilde{\nu}; \nuj, \alpha_j) = \sqrt{\frac{\ln 2}{\pi}}\frac{1}{\alpha_j}\exp\left( -\frac{(\tilde{\nu}-\nuj)^2 \ln 2}{\alpha_j^2} \right)
\end{align}
where the line centre position is $\nuj$ and the Doppler half-width at half-maximum,
\begin{align}
\alpha_j = \sqrt{\frac{2k_\mathrm{B}T\ln 2}{m}}\frac{\nuj}{c},
\end{align}
at temperature $T$ for a molecule of mass $m$.

\begin{figure}[h]
\begin{center}
\includegraphics[width=14cm]{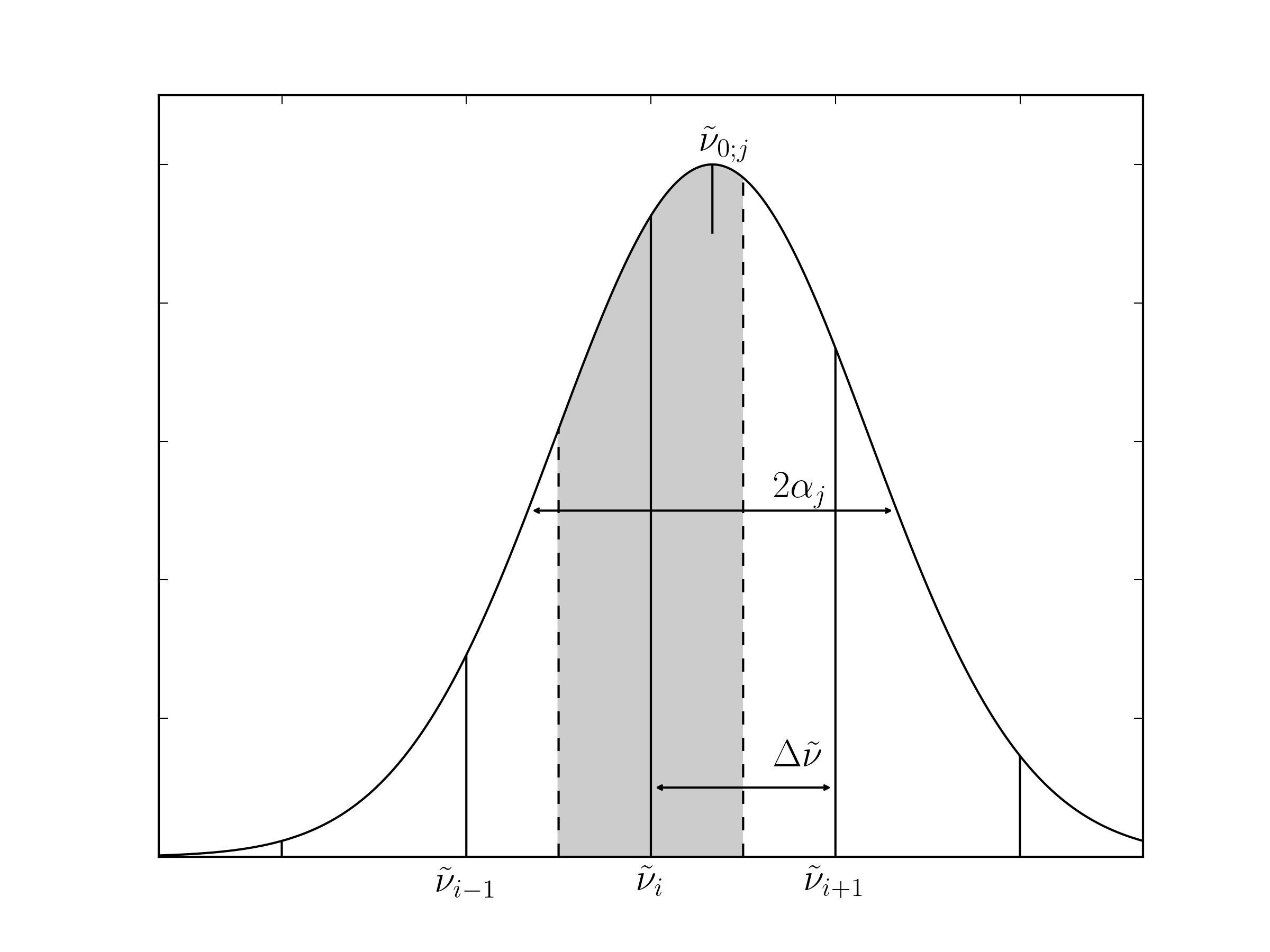}
\caption{The calculation of the absorption cross section in a wavenumber bin centered on $\nui$ due to a single line. The integrated line intensity within the shaded region, of width $\dnu$, contributes to $\sigma_{ij}$.}
\label{fig:line_fig}
\end{center}
\end{figure}

The contribution to the cross section within each bin is a sum over contributions from individual lines:
\begin{align}
\sigma_i = \sum_j \sigma_{ij}
\end{align}
where
\begin{align}
\label{eqn:sig_ij}
\sigma_{ij} &= \frac{S_j}{\dnu} \int_{\nui-\dnu/2}^{\nui+\dnu/2} f_\mathrm{G}(\tilde{\nu}; \nuj, \alpha_j) \,\mathrm{d}\tilde{\nu}\\
&= \frac{S_j}{2\dnu} \left[ \erf{x_{ij}^+} - \erf{x_{ij}^-} \right],
\end{align}

where erf is the error function and
\begin{align}
x_{ij}^\pm = \frac{\sqrt{\ln 2}}{\alpha_j}\left[ \nui \pm \frac{\dnu}{2} - \nuj \right]
\end{align}
are the scaled limits of the wavenumber bin centred on $\nui$ relative to the line centre, $\nuj$, and the line intensity in units of $\mathrm{cm^{-1}/(molecule\,cm^{-2})}$ is
\begin{align}
S_j = \frac{A_j}{8\pi c}\frac{g_j'\mathrm{e}^{-c_2E_j{''}/T}}{\nuj^2 Q(T)}\left( 1 - \mathrm{e}^{-c_2\nuj/T} \right).
\end{align}
Here, $g_j'$ and $E_j{''}$ are the upper-state degeneracy and lower-state energy respectively, $A_j$ is the Einstein $A$ coefficient for the transition and $c_2 \equiv hc/k_\mathrm{B}$ is the second radiation constant. For \htwoo, the molecular partition function, $Q(T)$, was obtained from the tabulated values of \citet{jt263}.

Note that in the limit of $\dnu \gg \alpha_j$, eqn (\ref{eqn:sig_ij}) reduces to
\begin{align}
\sigma_{ij} &\approx \frac{S_j}{\dnu} \int_{-\infty}^{+\infty} f_\mathrm{G}(\tilde{\nu}; \nuj, \alpha_j) \,\mathrm{d}\tilde{\nu} = \frac{S_j}{\dnu},
\end{align}
whereas for $\dnu \ll \alpha_j$,
\begin{align}
\sigma_{ij} &\approx S_j f_\mathrm{G}(\nui; \nuj, \alpha_j).
\end{align}
However, the exact expression in all calculations of the cross sections presented in this work.

\section{Results}
The absorption cross section of \htwoo\ was calculated between $10\;\cm$ and $30000\;\cm$ across the temperature range 296~K -- 3000~K (Table \ref{tab:T-summary}), using the wavenumber grid-spacing given in Table \ref{tab:dnu-summary}.

\begin{table}[tbp]
\caption{Temperatures at which calculated \htwoo\ cross sections are provided.}
\label{tab:T-summary}
\begin{center}
\begin{tabular}{rrrr}
\hline
296~K & 400~K & 500~K & 600~K\\
700~K & 800~K & 900~K & 1000~K\\
1200~K & 1300~K & 1400~K & 1600~K\\
1800~K & 2000~K & 2200~K & 2400~K\\
2600~K & 2800~K & 3000~K\\
\hline
\end{tabular}
\end{center}
\end{table}

\begin{table}[tbp]
\caption{Summary of the grid spacings, $\dnu$ for the cross sections calculated in different wavenumber regions}
\label{tab:dnu-summary}
\begin{center}
\begin{tabular}{ll}
\hline
wavenumber range $/\cm$ & $\dnu /\cm$\\
\hline
10 -- 100 & $10^{-5}$\\
100 -- 1000 & $10^{-4}$\\
1000 -- 10000 & $10^{-3}$\\
10000 -- 30000 & $10^{-2}$\\
\hline
\end{tabular}
\end{center}
\end{table}

\begin{figure}[h]
\begin{center}
\includegraphics[width=14cm]{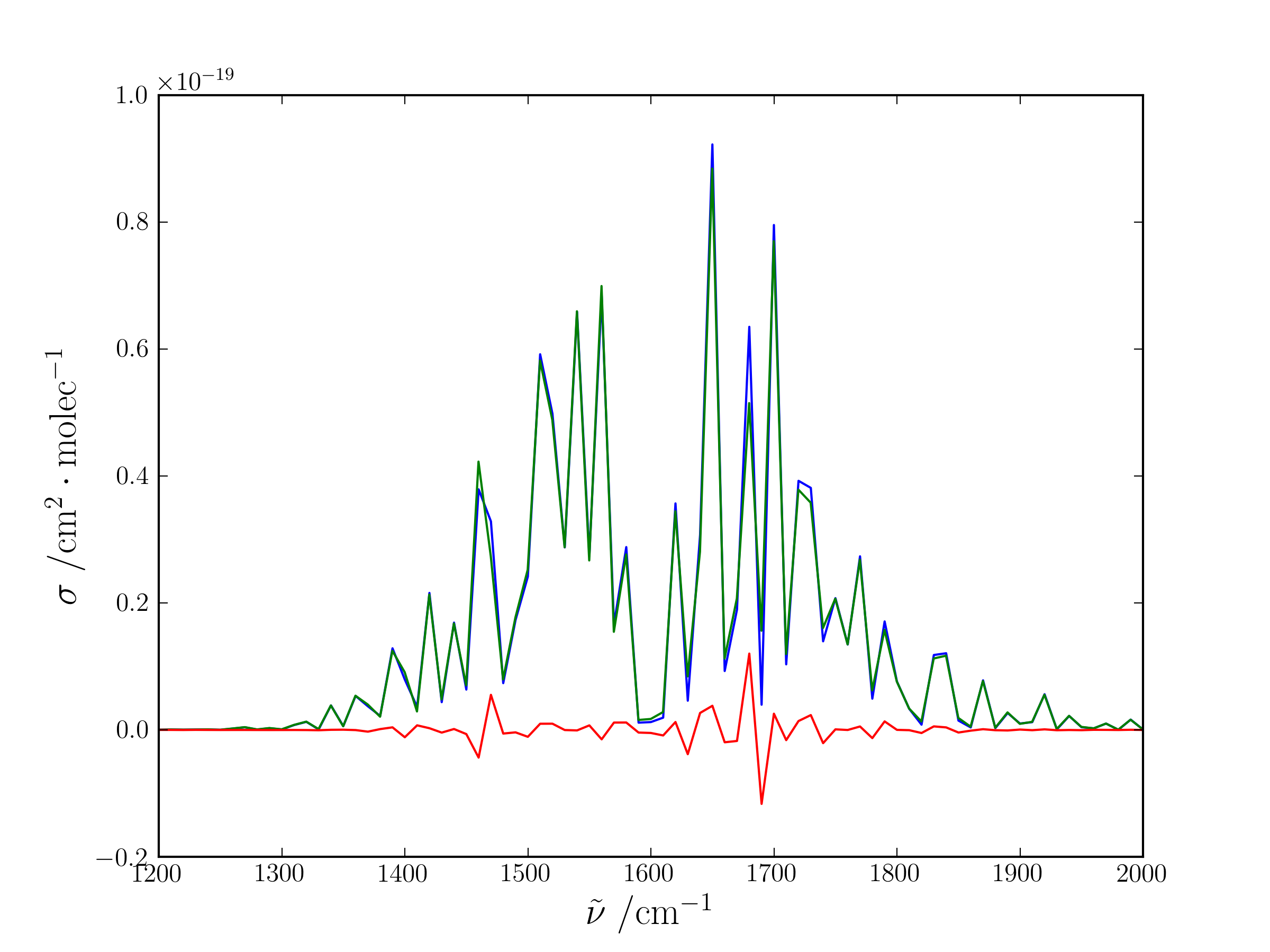}
\caption{Comparison of the calculated \htwoo\ cross section presented in this work (blue) with the experimental cross section of the PNNL database \citep{sjs04} (green) in the region of the fundamental $\nu_2$ bending mode, at 296~K, both binned to a $10\;\cm$ wavenumber grid. Also shown is the difference (this work - PNNL) between the two spectra (red).}
\label{fig:PNNL-comparison}
\end{center}
\end{figure}

For comparison with experimental spectra, low-resolution cross sections were produced by binning the high-resolution cross sections to the following fixed grid spacing across the entire wavenumber range: $\dnu = 0.01$, 0.1, 1, 10, $100\;\cm$. At these resolutions, the structure due to individual lines is lost and direct comparison can be made with, for example, the experimental water vapour cross sections of the PNNL database \citep{sjs04}. Such a comparison for the $\dnu = 10\;\cm$ resolution spectra is shown in Figure \ref{fig:PNNL-comparison}.

\section{Interpolation of cross sections between temperatures}

\begin{figure}[h]
\begin{center}
\includegraphics[width=14cm]{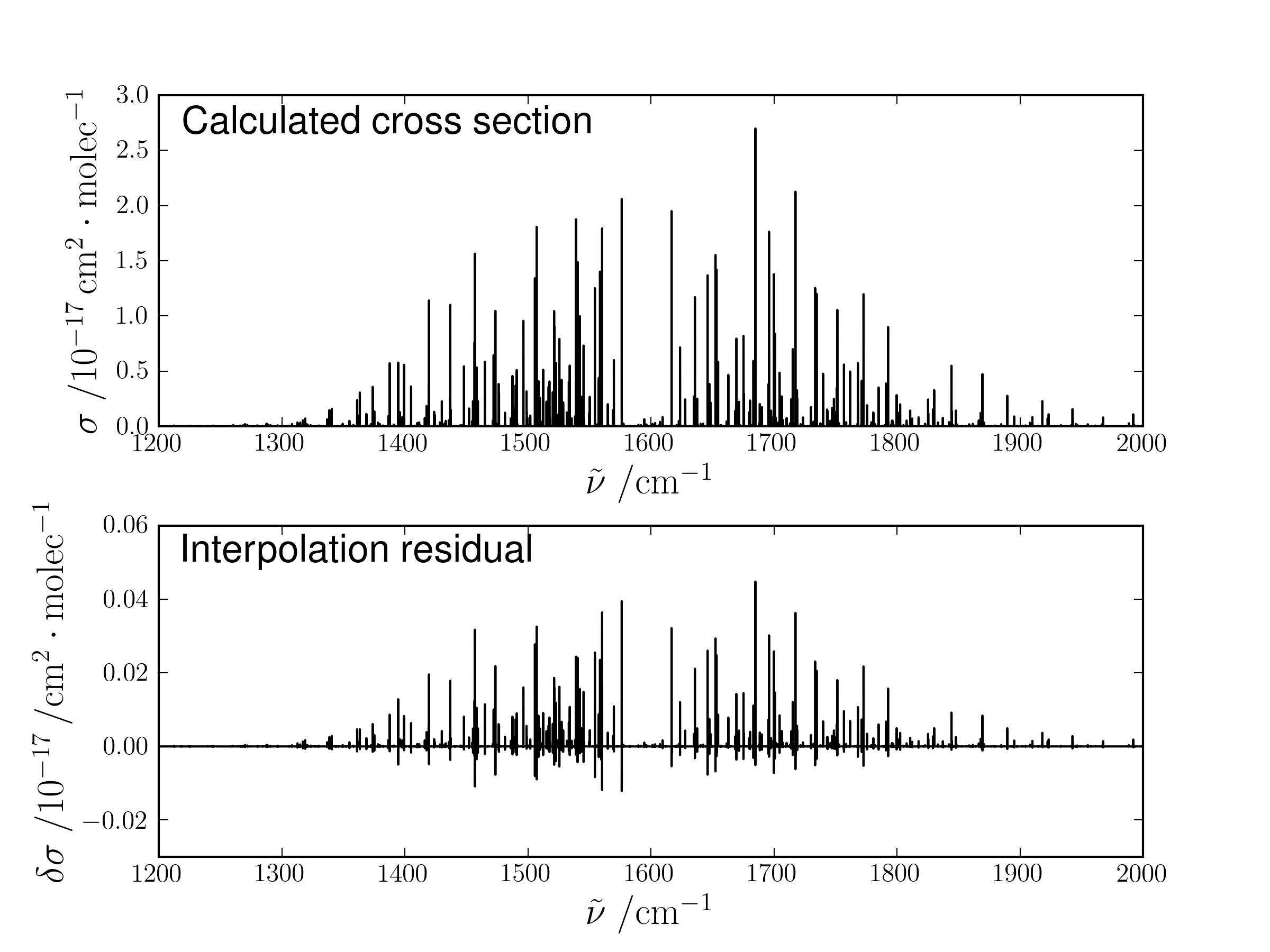}
\caption{Calculated absorption cross section (upper pane) and interpolation residual (lower pane) in the region of the fundamental $\nu_2$ bending mode, at 350~K, on a wavenumber grid spacing of 0.01~\cm. The interpolation residual error does not exceed 1.34\%.}
\label{fig:interp-error}
\end{center}
\end{figure}

For use in the web-based application described below, cross sections for the molecules given in Table \ref{tab:species} have been calculated using a wavenumber grid spacing of $0.01\;\mathrm{cm^{-1}}$ at a range of temperatures in 100~K intervals below 1000~K and 200~K intervals above 1000~K. A cross section at some intermediate temperature between the values at which the stored cross sections have been calculated may be obtained by interpolation. Suppose that $\sigma_i(T_1)$ and $\sigma_i(T_2)$ are calculated cross sections at temperatures which bracket the temperature of the desired cross section: $T_1 < T < T_2$ (we consider interpolation using only $\sigma_i$ calculated at the two temperatures closest to $T$). One possible interpolation strategy is the linear interpolation
\begin{align}
\sigma_i(T) = \sigma_i(T_1) + m(T-T_1), \; \mathrm{where} \; m = \frac{\sigma_i(T_2)-\sigma_i(T_1)}{T_2-T_1}.
\end{align}
However, we find a more accurate approach is to estimate the temperature dependence to be of the form
\begin{align}
\sigma_i(T) = a_i\mathrm{e}^{-b_i/T},
\end{align}
where the coefficients $a_i$ and $b_i$ at each wavenumber bin may be calculated from
\begin{align}
b_i = \left(\frac{1}{T_2}-\frac{1}{T_1}\right)^{-1}\ln \frac{\sigma_i(T_1)}{\sigma_i(T_2)}\;\;\;\mathrm{and} \;\;\; a_i = \sigma_i(T_1)\mathrm{e}^{b_i/T_1}.
\end{align}
The largest values of the interpolation residual error in the region 1000 - 20000 \cm, calculated as $\delta\sigma_i = \sigma_{i,\mathrm{calc}} - \sigma_{i,\mathrm{interp}}$, are found to be associated with the $\nu_2$ band - as an illustration, this is plotted in Figure \ref{fig:interp-error} at 350~K. The maximum value of the interpolation residual across this wavenumber region at a range of temperatures and wavenumber binning intervals is given in Table \ref{tab:interp-error}, expressed as a percentage of the corresponding absorption cross section:
\begin{align}
\delta\sigma_\mathrm{max}^\% = \max \left(\frac{|\sigma_{i,\mathrm{calc}} - \sigma_{i,\mathrm{interp}}|}{\sigma_{i,\mathrm{calc}}} \right) \times 100.
\end{align}
In all cases, $\delta\sigma_\mathrm{max}^\%$  is found to be less than the estimated uncertainty in the \emph{ab initio} line intensities that the cross section calculation is based on. Interpolation is performed on the 0.01~\cm~grid before binning to a coarser wavenumber grid, if required.

\begin{table}[tbp]
\caption{Maximum interpolation errors in the \htwoo\ cross section as a function of wave number grid spacing and temperature}
\label{tab:interp-error}
\begin{center}
\begin{tabular}{r|rrrr}
\hline
$\Delta\tilde{\nu}$ & 0.01 \cm & 0.1 \cm & 1.\cm & 10. \cm \\
\hline
350~K & 1.64 \% & 1.34 \% & 1.07 \% & 1.10 \% \\
1100~K & 1.00 \% & 0.50 \% & 0.49 \% & 0.46 \% \\
2500~K & 0.66 \% & 0.40 \% & 0.37 \% & 0.36 \% \\
\hline
\end{tabular}
\end{center}
\end{table}

Finally we note that \cite{12HaLiBe.NH3} recently presented an experimental ammonia spectrum
recorded at a range of temperatures at 100~K intervals. We suggest that our proposed interpolation
scheme would also be appropriate for interpolating their data.

\section{Web based application}

Calculated absorption cross sections can be obtained from the interface at the url http://www.exomol.com/xsecs. The user of this web-based interface can select a wavenumber range, temperature and wavenumber grid spacing; using these parameters the interface software first obtains a high-resolution cross section at the desired temperature by the interpolation procedure described in the previous section on the pre-calculated spectra, and then bins this interpolated cross section to the requested wavenumber grid.

Cross sections are returned as a list of floating point numbers in a text file, separated by the Unix-style newline character, \texttt{LF} (`\textbackslash n', \texttt{0x0A}). The wavenumber grid can be generated from the linear sequence
\begin{equation}
\tilde{\nu}_i = \tilde{\nu}_\mathrm{min} + i\Delta\tilde{\nu}; \quad i = 0,1,2,\cdots,n-1
\end{equation}
where the total number of points in the requested cross section is
\begin{equation}
n = \frac{\tilde{\nu}_\mathrm{max} - \tilde{\nu}_\mathrm{min}}{\Delta\tilde{\nu}} + 1.
\end{equation}
We also provide an XML file in XSAMS format \citep{xsams01}, compatible
with the standards of the VAMDC project \citep{jt481}. This file may
be thought of as a `wrapper' to the cross section data, providing
contextual metadata such as the molecular identity and structure,
temperature of the calculation, and wavenumber limits and grid
spacing. An example of the format is given in Table~\ref{tab:xsams-example}.

\begin{table}[tbp]
\caption{Sample XSAMS format \citep{xsams01} XML wrapper for a cross section for \htwoo\ generated from 1000 to 20000 \cm\
in steps of 1 \cm\ at a temperature of 296~K. In this example, the cross section itself is provided in the file \texttt{H2O\_1000-20000\_296K-10.0.sigma} as a column of values, here in $\mathrm{cm^2}$, one for each of 901 grid points.}
\label{tab:xsams-example}
\begin{verbatim}
<AbsorptionCrossSection envRef="EEXOMOL-1" id="PEXOMOL-XSC-1">
  <Description>
  The absorption cross section for H2O at 296.0 K, calculated at
  Sun Mar 11 19:50:45 2012, retrieved from www.exomol.com/xsecs
  </Description>
  <X parameter="nu" units="1/cm">
    <LinearSequence count="901" initial="1000." increment="10."/>
  </X>
  <Y parameter="sigma" units="cm2">
    <DataFile>H2O_1000-20000_296K-10.0.sigma</DataFile>
  </Y>
  <Species>
    <SpeciesRef>XEXOMOL-XLYOFNOQVPJJNP-FNDQEIABSA-N</SpeciesRef>
  </Species>
</AbsorptionCrossSection>
\end{verbatim}
\end{table}

\begin{table}[tbp]
\caption{Summary of species for which are cross sections currently available.
Also given for each species is the maximum wavenumber ($\ensuremath{\tilde{\nu}_{\mathrm{max}}}$), the maximum temperature ($T_{\mathrm{max}}$) and the reference to the original
line list.}
\label{tab:species}
\begin{center}
\begin{tabular}{lccl}
\hline
Species&  $\ensuremath{\tilde{\nu}_{\mathrm{max}}}$/\cm & $T_{\mathrm{max}}$/K& Reference\\
\hline
H$_3^+$ & 10~000 & 4000 & \citet{jt181}\\
H$_2$D$^+$ & 10~000 & 4000 & \citet{jt478}\\
H$_2$O & 20~000 & 3000 & \citet{jt378}\\
HDO & 17~000 & 3000 & \citet{jt469}\\
HCN / HNC  &  10~000 & 4000 & \citet{jt298,jt374}\\
H$^{13}$CN / H$^{13}$NC&  10~000 & 4000 & \citet{jt447}\\
NH$_3$ &  12~000 & 1500 & \citet{jt500}\\
\hline
\end{tabular}
\end{center}
\end{table}

Cross section files have been generated for the polyatomic line lists
currently available on the ExoMol website. These are listed in
Table~\ref{tab:species}. The table also specifies the maximum
wavenumber ($\ensuremath{\tilde{\nu}_{\mathrm{max}}}$) and maximum
temperature ($T_{\mathrm{max}}$) for each species; we strongly caution against relying on
the cross sections or indeed the underlying line lists at temperatures greater than those given.
Further cross sections will be provided as line lists for new species  as they
become available.

\section{Conclusion}

High resolution absorption cross sections have been calculated for a
number of molecules likely to be important in the atmospheres of
exoplanets. The online interface provided at the ExoMol website
(www.exomol.com) allows customized cross sections for a given
molecular species to be returned at a specified temperature and
resolution. Cross sections are only available for those species for
which extensive line lists exist. New cross sections will
be provided as further species are added to the ExoMol database,
see  \citet{jt529} for example.

It is our intention to make the cross section facility in ExoMol
fully interoperable with other  spectroscopic
databases as part of the VAMDC (Virtual Atomic and Molecular Data Centre) project
\citep{jt481}. Work in this direction will be reported in due course.

\section*{Acknowledgements}
We thank Giovanna Tinetti and Bob Barber
for many helpful discussions during the course of this work.
This work was performed as part of ERC Advanced Investigator Project 267219
and the project VAMDC which is funded by the European Union INFRA-2008-1.2.2 Scientific Data Infrastructure program under
Grant Agreement number 239108.



\bibliographystyle{elsarticle-harv}


\end{document}